# Bayesian Population Receptive Field Modelling

Peter Zeidman[a] *

Edward Harry Silson[b]

Dietrich Samuel Schwarzkopf[c,d]

Chris Ian Baker[b]

Will Penny[a]

[a] The Wellcome Trust Centre for Neuroimaging, University College London, 12 Queen Square, London, UK. WC1N 3BG.

[b] Laboratory of Brain and Cognition, National Institute of Mental Health, Bethesda, Maryland 20892-1366, USA.

[c] Experimental Psychology, University College London, 26 Bedford Way, London, UK. WC1H 0AP.

[d] UCL Institute of Cognitive Neuroscience, 17-19 Queen Square, London, UK. WC1N 3AR.

* Corresponding author

peter.zeidman@ucl.ac.uk

The Wellcome Trust Centre for Neuroimaging is supported by core funding from the Wellcome Trust 091593/Z/10/Z.

**Abstract**

We introduce a probabilistic (Bayesian) framework and associated software toolbox for mapping population receptive fields (pRFs) based on fMRI data. This generic approach is intended to work with stimuli of any dimension and is demonstrated and validated in the context of 2D retinotopic mapping. The framework enables the experimenter to specify generative (encoding) models of fMRI timeseries, in which experimental manipulations enter a pRF model of neural activity, which in turns drives a nonlinear model of neurovascular coupling and Blood Oxygenation Level Dependent (BOLD) response. The neuronal and haemodynamic parameters are estimated together on a voxel-by-voxel or region-of-interest basis using a Bayesian estimation algorithm (variational Laplace). This offers several novel contributions to receptive field modelling. The variance / covariance of parameters are estimated, enabling receptive fields to be plotted while properly representing uncertainty about pRF size and location. Variability in the haemodynamic response across the brain is accounted for. Furthermore, the framework introduces formal hypothesis testing to pRF analysis, enabling competing models to be evaluated based on their model evidence (approximated by the variational free energy), which represents the optimal tradeoff between accuracy and complexity. Using simulations and empirical data, we found that parameters typically used to represent pRF size and neuronal scaling are strongly correlated, which should be taken into account when making inferences. We used the framework to compare the evidence for six variants of pRF model using 7T functional MRI data and we found a circular Difference of Gaussians (DoG) model to be the best explanation for our data overall. We hope this framework will prove useful for mapping stimulus spaces with any number of dimensions onto the anatomy of the brain.

# 1 Introduction

There are many examples of neuronal populations which represent stimulus spaces. In the auditory cortex, the 1-dimensional space of sound frequencies is mapped onto the surface of the brain (Merzenich and Brugge, 1973; Moerel et al., 2012). In the visual system, retinotopic mapping has revealed that the 2-dimensional plane of the retina is mapped multiple times onto the surface of visual cortex (e.g. Holmes, 1945). Place cells in the bat hippocampus respond maximally to a specific location in 3-dimensional space (Palacci et al., 2013) and conceptual knowledge may be represented neuronally in spaces of two dimensions or more (Constantinescu et al., 2016). Populations of neurons can be characterised by their receptive fields – the area(s) of N-dimensional space to which they maximally respond. In this paper, we introduce a generic framework for mapping stimulus spaces onto the brain and for performing hypothesis testing. We illustrate this approach in the context of visual population receptive field (pRF) mapping.

To enable pRF mapping, model parameters are required which capture the response of neuronal populations to experimental stimuli. The spatial distribution of these parameters across the brain can reveal large-scale topographic features, such as the presence of retinotopic maps. pRF mapping can therefore be seen a special case of building generative models of imaging timeseries. We seek to understand how stimuli cause a change in neuronal activity, which in turns causes a change in observations. For functional MRI (fMRI), this involves modelling neuro-vascular coupling and the BOLD response (Kumar et al., 2014), which is an inherently nonlinear mapping. For instance, the BOLD response has a refactory period which depends on the interstimulus interval (Friston et al., 1998). Furthermore, brain regions differ in the extent of their vascularization, giving rise to regional differences in BOLD response. Typically, pRF mapping experiments use a canonical haemodynamic response function, which may be determined on a per-subject basis. Here, to obtain the best possible estimators of neural activity for constructing pRF maps, we specified and estimated a non-linear model for each voxel's fMRI timeseries which included a biologically motivated differential equation model of neurovascular coupling and the BOLD response (Buxton et al., 2004; Stephan et al., 2007).

The objective of modelling (and of science more generally) is to test hypotheses. In the context of pRF mapping, hypotheses may be specified explicitly or implicitly. For instance, He et al. (2015) tested the explicit hypothesis that pRF position is modulated by perceived 3D space. Other pRF studies have been exploratory, for instance examining the reorganisation of visual field maps after lesions or disease (e.g. Levin et al., 2010). In studies such as this, there is an implicit hypothesis that pRF parameters will deviate from the normal range in specific areas of cortex. Despite the popularity of pRF mapping, a framework for formal hypothesis testing is currently lacking.

Here we introduce a set of tools for probabilistic (Bayesian) model fitting and inference in the context of pRF mapping, which could offer several benefits to experimenters. The optimal method for testing hypotheses is to compare the likelihood of the data under one model (or hypothesis) against the likelihood of the data under another model (Neyman and Pearson, 1933). For instance, an experimenter may wish to test the hypothesis that in certain brain regions there are receptive fields with an excitatory centre and inhibitory surround, as identified by Hubel and Wiesel (1959) with single unit recordings. Such receptive fields may be modelled using a Difference of Gaussians (DoG) function (Rodieck, 1965), which can also capture the neuronal response at the level of voxels in fMRI data (Zuiderbaan et al., 2012). Alternatively, if the evidence for an inhibitory surround is lacking, a simple excitatory receptive field may be the better model (as applied to fMRI data by Dumoulin and Wandell, 2008). This kind of question, regarding which of several models is the best explanation of the available imaging data, may be addressed by comparing the evidence for the fMRI data under competing models at each point in the brain.

Normally, pRF models are assessed based on the percentage variance they explain (their accuracy). This cannot be used for hypothesis testing, as it ignores complexity – any model with more (independent) parameters will explain more of the variance, with the added risk of overfitting the noise and failing to generalise. In the framework proposed here, an approximation of the model evidence is used known as the negative variational free energy (Friston et al., 2007; Penny, 2012). This quantity, estimated for each pRF model, represents the accuracy of the model minus its complexity. By comparing models based on their free energy, the experimenter can select the simplest model which explains the most variance. Furthermore, by taking into account the covariance between parameters, the free energy offers a more sensitive approximation to the model evidence than other approximations such as the AIC or BIC.

As well as enabling competing models to be compared, the framework we propose has advantages for parameter-level inference, which may be of particular relevance for exploratory pRF studies. Here, parameters such as the pRF's size are each represented as a (normal) probability distribution, with both an expected value and variance / covariance (uncertainty). Thus, the uncertainty of parameter estimates may be expressed when visualising the pRF and when making comparisons within and between subjects. Uncertainty about the parameter estimates may arise from multiple sources – observation noise, subjects' movement, as well as any covariance among parameters. Also, it may not always be possible to confidently assign variance in the measured signal to either neuronal or haemodynamic causes. By estimating the full covariance among neuronal and haemodynamic parameters, we ensure that any uncertainty induced by ambiguity between these parameters is accounted for when visualising the pRF or testing hypotheses.

Here, we generalise an approach previously introduced in the context of tonotopic mapping (Kumar et al., 2014), making several novel contributions. We extend the method to stimuli of any dimension, and demonstrate its application in the context of visual (retinotopic) pRF mapping (Section 3.1, 3.2). We evaluate the face validity and robustness to noise of the method using simulated data (Section 3.3), and evaluate test-retest reliability across scanning runs using empirical data (Section 3.4). Finally, we demonstrate the use of this method for hypothesis testing (Section 3.5), by comparing the evidence for two established forms of pRF model: a Gaussian response function (Dumoulin and Wandell, 2008) and a Difference of Gaussians (centre-surround) response function (Zuiderbaan et al., 2012). Within each category of model we also compared the evidence for circular, elliptical, or angled receptive fields. All of the methods described and evaluated here are made available to experimenters via a freely available software toolbox (Appendix A).

## 2 Methods

### 2.1 Participants

Empirical data were acquired as part of a previously reported study (Silson et al., 2015). Data from one participant was included here. All participants in the previous study had normal or corrected-to-normal vision and gave written informed consent. The National Institutes of Health Institutional Review Board approved the consent and protocol (93-M-0170, NCT00001360).

### 2.2 Data acquisition

Data were acquired using a Siemens 7 tesla Magnetom scanner in the Clinical Research Center on the National Institutes of Health campus (Bethesda, MD). Partial EPI volumes of the occipital and temporal cortices were acquired using a 32-channel head coil (42 slices; 1.2 x 1.2 x 1.2 mm; 10% interslice gap; TR, 2 s; TE, 27 ms; matrix size, 170 x 170; FOV, 192 mm). Anatomical T1 weighted volumes were acquired before the experimental runs. Standard MPRAGE (Magnetization-Prepared Rapid-Acquisition Gradient Echo) and corresponding GE-PD (Gradient Echo–Proton Density) images were collected and the MPRAGE images were then normalized by the GE-PD images, for use as a high-resolution anatomical data for the fMRI data analysis.

## 2.3   Task and procedure

Naturalistic scene images were presented through a bar aperture that gradually traversed the visual field (Figure 1). During each 36 s sweep, the aperture took 18 evenly spaced steps (each 2s or 1TR) to traverse the entire screen (Dumoulin and Wandell, 2008). Eight of these sweeps formed one run, in the following order: L-R, BR-TL, T-B, BL-TR, R-L, TL-BR, B-T, and TR-BL. There were 16 identical runs per participant. The scene stimuli, which covered a circular area (21° diameter) changed every 400ms (5 per aperture position). During runs, participants performed a colour-detection task at fixation, indicating via button press when the white fixation dot changed to red. Colour fixation changes occurred semi-randomly, with ~2 colour changes per sweep.

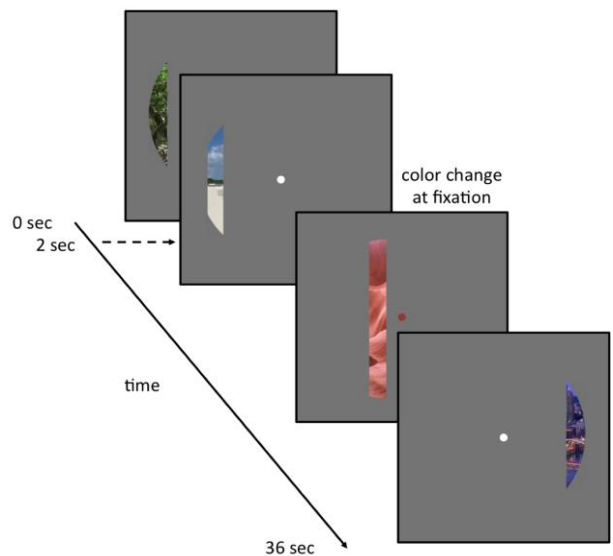

*Figure 1* Example stimulus frames used for pRF mapping. Scene images covering a circular (21° diameter) area of the display were presented through a bar aperture that moved gradually through the visual field. A single sweep across the visual field took 36 s and consisted of 18 equally timed (2 s) steps of equal width. In each run, the aperture completed eight sweeps (2 orientations, 4 directions). Participants were required to maintain fixation and indicate the detection of a colour change at fixation, via a button press. Adapted from Silson et al. 2015 (license CC BY 4.0).

## 2.4   Pre-processing

As reported in Silson et al. (2015), functional data were initially motion-corrected using the Analysis of Functional NeuroImages (ANFI) software package (http://afni.nimh.nih.gov/afni), following removal of eight "dummy" volumes to allow stabilisation of the magnetic field. Subsequent analyses were conducted using SPM (Statistical Parametric Mapping) version 12 (http://www.fil.ion.ucl.ac.uk/spm/).

## 2.5   Feature selection

To reduce computation time of the Bayesian pRF procedure we performed an initial General Linear Model (GLM) analysis to pick out any voxels showing some differential response depending on the location of the bar aperture. This also enabled us to model and regress out nuisance effects (e.g. motion and scanner drift). We first divided the stimulus display into 9 equally sized squares. For each 2s frame of the stimulus, we identified which of the 9 squares had any illumination. We then built a GLM with one regressor for each square, as well as 6 regressors representing head motion and a regressor modelling the mean of each run (Figure 2). We estimated the parameters of this model and created an F-contrast to identify positive or negative effects from any of the first 9 regressors (defined as an identity matrix of dimension 9). The resulting statistical image, thresholded at p < 0.05 family-wise error corrected (F = 5.48) revealed 14,395 voxels located primarily in the occipital lobe, extending ventrally into the posterior fusiform / parahippocampal region. These voxels were taken forward for pRF analysis.

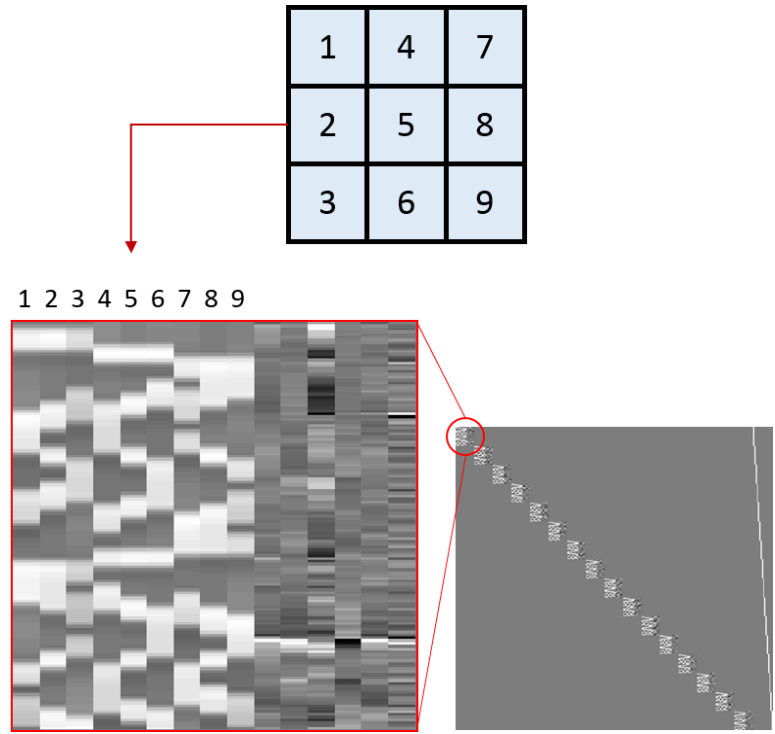

*Figure 2* General linear model for feature selection. We split the stimulus space into 9 squares (top) and constructed a general linear model (bottom right). One run out of 16 is enlarged (bottom left). The first 9 columns represent locations in stimulus space, the remaining 6 columns represent the subject's head motion.

## 2.6 Timeseries extraction

pRF modelling was performed on a per-voxel basis. To enable this, timeseries were extracted from each voxel identified above, high-pass filtered (with cut-off 128 seconds) to remove low frequency scanner drift and whitened to remove auto-correlation over time. The motion regressors and mean of each run were regressed out. This procedure was repeated for each scanning run and the resulting timeseries were averaged over runs. Data from the 8 odd-numbered runs were used for the main analysis and the 8 even-numbered runs were kept aside to evaluate test-retest reliability (Section 3.4).

## 2.7 Model specification

We specified a generative model which predicted BOLD timeseries given visual stimuli. The model (based on Kumar et al., 2014) consisted of two parts – a neuronal model describing the brain's response to the stimuli, and an observation model describing the change in neurovascular coupling and BOLD response caused by the neuronal activity.

### 2.7.1 Stimulus specification

At each time step $t$, a set of points $U_t$ was illuminated on the screen, where each point was defined by an $x$- and $y$-coordinate:

$$U_t = \{p_1, \ldots, p_n\}$$

$$p_i = (x \in \mathbb{R}, y \in \mathbb{R})$$

(1

The on-screen coordinates were discretised to give a reduced stimulus resolution of 41x41 pixels and expressed in units of degrees of visual angle.

### 2.7.2 Neuronal model

The neuronal activity $z(t)$ was modelled as a multivariate normal probability density function $N$:

$$z(t) = \beta \sum_{p_i \in U_t} N(p_i | \mu, \Sigma)$$

(2

$$N(p_i | \mu, \Sigma) = \frac{1}{\sqrt{(2\pi)^k |\Sigma|}} e^{-\frac{1}{2}(p_i - \mu)' \Sigma^{-1}(p_i - \mu)}$$

This function is commonly used in pRF mapping (since Dumoulin and Wandell, 2008), although we used a normalised rather than the unnormalised density $N$. The response was summed over each illuminated pixel on the screen $p_i$ and scaled by parameter $\beta$. The model was parameterised by $k$-dimensional vector $\mu$, which specified the position of the pRF in $k$-dimensional stimulus space, and $k$-dimensional covariance matrix $\Sigma$ which specified the width and rotation of the receptive field. Here we used stimuli of dimension $k = 2$, specified by their Cartesian coordinates $(x, y)$ in visual space.

The neuronal parameters ($\mu$, $\beta$ and the elements of matrix $\Sigma$) needed certain constraints to be placed on their values. For instance, $\beta$ needed to be positive whereas $\mu$ was constrained within the stimulated area of the visual field. However, the optimisation algorithm we employed (variational Bayes) expected normally distributed prior distributions for each parameter, constrained only by their variance. To overcome this, we substituted each parameter for a latent variable, which could be freely adjusted by the optimisation algorithm. Within the pRF model, these variables were transformed to be within the desired range of values, before being used to calculate the neuronal response in Equation 2. Next, we detail these transformations for each parameter.

**pRF centre parameters**

We constrained the centre of the pRF to fall within the circular area of the screen that was illuminated during scanning. While the pRF approach can model the influence of stimulation from beyond the stimulated area, finding the parameters for such a model is a particularly ill-posed problem (any number of parameter combinations could give rise to the same observations). To constrain the pRF location we introduced latent variables $l_\rho$ and $l_\theta$, which controlled the distance and angle (polar coordinates) of the pRF location relative to the centre of the visual field. These variables were freely adjusted by the optimisation algorithm during model fitting, and were transformed to polar coordinates $(\rho, \theta)$ within the model as follows:

$$\rho = r * \text{NCDF}(l_\rho, 0, 1)$$

(3

$$\theta = 2\pi * \text{NCDF}(l_\theta, 0, 1) - \pi$$

Where NCDF is the cumulative density function for the univariate normal distribution. These transforms ensured that the distance $\rho$ of the pRF from the centre of the visual field was constrained between 0 and $r = 10.5°$ and the polar angle $\theta$ was constrained between $-\pi$ and $\pi$ radians. A second transform then converted from polar to Cartesian coordinates:

$$\mu_x = \rho * \cos(\theta)$$

$$\mu_y = \rho * \sin(\theta) \quad (4$$

$$\mu = (\mu_x, \mu_y)$$

Which then entered the model as $\mu$ in Equation 2. We could therefore guarantee that whatever values were selected for latent variables $l_\rho$ and $l_\theta$ by the optimisation algorithm, the pRF location $(\mu_x, \mu_y)$ would fall within the circular stimulated area.

**pRF width parameter**

The width of the pRF was parameterised by its standard deviation $\sigma$. To constrain the width to be in the range $[r_0\ r]$, we introduced a latent variable $l_\sigma$. This latent variable could be freely adjusted by the optimisation algorithm and was transformed within the model to give the pRF width:

$$\sigma = (r - r_0) * \text{NCDF}(l_\sigma, 0, 1) + r_0 \quad (5$$

We set $r_0$, the smallest allowed pRF width, to $0.5°$. The pRF model which forms the first part of the results (Section 3.1-3.4) had a receptive field which was circular in shape (isotropic). This was modelled by setting the covariance matrix $\Sigma$ in Equation 2 as follows:

$$\Sigma = \begin{bmatrix} \sigma^2 & 0 \\ 0 & \sigma^2 \end{bmatrix} \quad (6$$

**Scaling parameter**

Parameter $\beta$ scaled the neuronal response (see Equation 2). It was constrained to be positive by substituting it for latent variable $l_\beta$ :

$$\beta = \exp(l_\beta) \quad (7$$

**Priors**

To complete the specification of the neuronal model, each latent variable $(l_\rho, l_\theta, l_\beta)$ was assigned a prior (normal) distribution, representing our beliefs before model fitting. The latent variables representing the pRF location $l_\rho$ and $l_\theta$ had priors $N(0,1)$. The transformation to constrain these (Equation 3) converted these normal prior distributions to bounded flat distributions over $\rho$ and $\theta$, with all values in the allowed ranges equally likely (this is known as the probability integral transform). The second transformation, from polar to Cartesian coordinates (Equation 4), changed this to a peaked distribution around the origin. We considered this a reasonable prior given the over-representation of the fovea in cortex (e.g. Azzopardi and Cowey, 1993).

We set the prior on the latent variable representing the pRF width $l_\sigma$ to $N(0,1)$. Because of the CDF transformation (Equation 5), this gave an equal prior probability of all widths in the allowed range. We set $l_\beta$ to have prior distribution $N(-2,5)$, which translated to an expected $\beta$ parameter of 0.14 with 95% confidence interval [0.003 5.35]. This large variance reflected our uncertainty about this parameter. We found that having a small positive prior value for this parameter provided a reasonable starting point for the optimisation algorithm.

### 2.7.3 Haemodynamic model

The predicted neuronal activity from the pRF model was fed into the extended Balloon model (Buxton et al., 2004; Stephan et al., 2007), which is a series of differential equations describing how neuronal activity causes a change in blood flow, and how blood flow causes the BOLD response. Most parameters of this model are fixed, based on previous empirical measurements, however three parameters are estimated on a voxel-wise basis: the transit time $\tau$, the rate of signal decay $\kappa$, and the ratio of intra- to extra-vascular signal $\epsilon$. Certain parameters in this model are field-strength specific, which we adjusted for our 7T data according to recommendations by Heinzle et al. (2016). We set $r_0$, the intravascular relaxation rate, to 340, the frequency offset at the outer surface of magnetized vessels to 197.86Hz and the prior distribution on $\log \epsilon$ to the normal distribution with mean -3.99 and variance 0.83. The 3 free neurovascular / haemodynamic parameters were concatenated with the parameters of the neuronal model and estimated simultaneously.

The model was completed by setting prior expectations on the observation noise, which was assumed to be I.I.D. and normally distributed with mean zero. We set the prior log precision of the noise to 4 with variance (uncertainty) 1/55, based on values typically used with SPM's Bayesian modelling framework.

## 2.8 Model estimation

For each voxel separately, the model was fitted to the data using the variational Laplace algorithm (Friston et al., 2007), implemented in the Matlab function spm_nlsi_gn.m, which is included with the SPM software package. This Bayesian estimation procedure provides two important estimates. The first is an approximation of the log model evidence, which is the log probability of the observing the data $y$ given the model $m$, $\log p(y|m)$. The approximation, known as the negative variational free energy (henceforth 'free energy'), may be expressed as follows:

$$\log p(y|m) = F(m) + KL[q(\theta|m) \;||\; p(\theta|y,m)] \quad (8$$

Here, $F$ is the free energy approximation of the model evidence, which we estimate. The second term is a distance measure (the Kullback-Leibler divergence, KL) between two distributions: the estimated parameters $q(\theta|m)$ and the true parameters $p(\theta|y,m)$. We cannot compute this, as we do not know the true parameters. However, because the log model evidence $\log p(y|m)$ is fixed, we know that by maximising the free energy $F(m)$, we minimise the second term and get the closest approximation of the model evidence. The free energy is defined in full elsewhere (see Appendix B of Penny, 2012), however it is important to recapitulate that it may be decomposed into two parts (Beal and Ghahramani, 2003):

$$F(m) = \text{accuracy}(m) - \text{complexity}(m) \quad (9$$

The accuracy is the fit of the model to the data. The complexity is the difference (the KL divergence) between the prior distribution of the parameters and the estimated (posterior) distribution of the parameters. This definition of complexity gives the free energy advantages over other approximations such as the AIC and BIC (Penny, 2012). Rather than simply counting the number of parameters in the model, the KL divergence takes into account the full covariance among parameters, meaning that parameters which are estimated to be independent contribute more to the complexity term than those which covary. Using the free energy, we compared pRF models to find the most accurate and least complex explanation for our data.

The algorithm also provides the estimated parameters $q(\theta|m)$ which maximise the free energy. The algorithm employs the Laplace assumption, which means that all parameters – prior and posterior – are normally distributed. Thus for each parameter, whether neuronal or haemodynamic, we have an expected value (mean) and uncertainty (variance), as well as the estimated covariance among parameters. We report these parameters and their uncertainty in various forms throughout the paper.

We performed model estimation separately for timeseries from each voxel. To improve performance, we initialized the VL model fitting algorithm with pRF parameters identified using a grid search (Dumoulin and Wandell, 2008) with a canonical haemodynamic response function. To further reduce the total estimation time, we developed software functions to divide the voxels across a parallel computer cluster and combine the results following estimation, which are included in the software implementation (Appendix A). Here we divided the estimation across a 192-core cluster computer (running CentOS 6.3 with a clock speed ranging from 2.1Ghz to 2.5Ghz). On each core of this cluster, estimation took an average of 123 seconds per voxel to complete, or around 2 hours 40 minutes for the whole analysis.

## 2.9 Inference

### 2.9.1 Parameter map generation

Having estimated the pRF models in each voxel, we generated maps of the estimated pRF parameters across the brain (Section 3.1). We then thresholded these parametric maps to only include voxels where something was learnt about the parameters after seeing the data, relative to the priors. To perform this thresholding, we compared the evidence for each estimated pRF model against a nested model, in which the pRF parameters were fixed at their prior means. If the evidence for these two models was similar, it meant that our knowledge about the pRF parameters did not improve as a result of seeing the data. On the other hand, if the evidence for the full model was stronger than the evidence for the nested model, it would mean that we can be more confident about these parameters after seeing the data.

To compute this model comparison, for each voxel's pRF model we specified a nested model $m_N$, in which the prior variance of the pRF location and width parameters ($\mu_x, \mu_y, \sigma$) were fixed at zero. We then computed the ratio of the evidence of the full model $m_F$ against this nested model $m_N$. This ratio is called the Bayes factor, and the log of this ratio is approximately equal to the difference in the free energy of each model:

$$\log \text{BF} = \log \frac{p(y|m_N)}{p(y|m_F)} \cong F_N - F_F \qquad (10$$

To compute the free energy of the nested models, $F_N$, we did not need to separately estimate them using the VL algorithm. Instead, we derived the nested models' evidence and parameters analytically using Bayesian Model Reduction (Friston and Penny, 2011), which is a generalisation of the Savage-Dickey ratio used in classical statistics. This calculation was performed on the order of milliseconds for each voxel. Having computed the log Bayes factor for each voxel, we then computed the posterior probability of the full model $m_F$:

$$p(m_N|y) = \frac{1}{1 + \exp(-\log \text{BF})}$$

$$p(m_F|y) = 1 - p(m_N|y)$$

(11

We thresholded the parametric maps to only include voxels where the posterior probability of the full model was at least 0.95. These surviving voxels were used to generate the maps in Figure 3.

### 2.9.2 Plotting haemodynamic response

To illustrate the estimated variability in BOLD response across the brain, we computed the impulse response function of each pRF model (those with significant pRF parameters, above) and averaged these responses across voxels (Section 3.1). The response functions were computed using Volterra kernels, as described by Friston et al. (1998).

### 2.9.3 Plotting pRFs

We plotted the receptive field of an example pRF model before and after fitting it to the data (Figure 5). The prior receptive field was computed by taking 1000 samples from model's prior multivariate distribution over the parameters, and for each sample, computing the response of the neuronal function (Equation 2) at evenly spaced locations in the visual field (Figure 5A). These responses were then averaged. This is referred to as the prior predictive density (PD) $q$ for each point in space $(x, y)$:

$$q(x, y) = \int g(x, y, \theta) p(\theta) d\theta$$

(12

Where $g(x, y, \theta)$ is the response of the neuronal (pRF) function to a single point in space with parameters $\theta$. By using this approach, we were able to plot the pRF while accounting for uncertainty over its location and size. The posterior response was similarly computed using the estimated parameters following model fitting.

### 2.9.4 Bayesian Parameter Averaging

In order to summarise pRF models' expected values and covariance across voxels, we computed the Bayesian Parameter Average (BPA). The BPA is the average of the parameters across voxels, weighted by the precision of each estimate. The data entering BPA was the vector of estimated pRF parameters $\mu_v$ for each voxel $v$, together with the precision matrix $\Lambda_v$ (inverse of the covariance matrix). The BPA provides a probability distribution over the parameters, with vector of means M and covariance matrix $C$:

$$\Lambda = \sum_{v=1}^{N} \Lambda_v$$
$$C = \Lambda^{-1} \tag{13}$$
$$M = \Lambda^{-1} \sum_{v=1}^{N} \mu_v \Lambda_v$$
$$p(M_i > 0) = 1 - NCDF(0; |M_i|, C_{i,i})$$

The final line of equation 13 states the probability that any given parameter $i$, averaged across voxels, is different to zero. $NCDF(x; a, b)$ is the normal cumulative distribution function with mean $a$ and variance $b$. In this formulation, we made the simplifying assumption that each voxel's data were independent.

### 2.9.5 Computing correlations and entropy maps

To compute the correlation between parameters we converted the covariance matrix $C$ from the BPA (described above) to a correlation matrix in the normal way. To visualise the spatial distribution of parameter uncertainty, we created maps of the entropy, defined for each voxel $v$ as:

$$h_v = \log |\Sigma_v| \tag{14}$$

Where $|...|$ is the determinant and $\Sigma_v$ is the estimated covariance matrix of the model for voxel $v$. The entropy $h_v$ of a multivariate Gaussian has units of nats, where more positive values indicate greater uncertainty in the parameters.

### 2.9.6 Simulation

To evaluate the face validity of the approach, we performed simulations. First, we estimated the empirical signal-to-noise ratio (SNR) of our data. To do this we randomly selected an estimated pRF model, which had been fitted to the timeseries from a voxel in right hemisphere near the midline (within the hOC2 anatomical mask, described in section 2.9.8). We integrated this model over time to give the predicted BOLD timeseries $y$ and computed the residual timeseries $r$. We calculated the SNR based on the standard deviation:

$$SNR = \frac{\sigma_y}{\sigma_r} = 1.56 \tag{15}$$

Next we chose a random 1000 voxels from our empirical data (which had survived the test for significance described in section 2.9.1), and we used the estimated pRFs from these voxels to generate 1000 simulated BOLD timeseries. We then added five levels of observation noise to each timeseries, from very noisy (SNR 0.5) down to a level of noise similar to our empirical data at 7T (SNR 1.56). We fitted pRF models to these simulated timeseries and correlated the estimated parameters against the 'true' parameters that generated the data. Note that for all validation analyses, we converted from latent variables (e.g. $l_\rho$ and $l_\theta$) to the underlying parameters which entered the model (e.g. $\rho$ and $\theta$), as these are the quantities of interest to experimenters.

### 2.9.7 Alternative models

We compared pRF models to address two questions. The first was whether a single Gaussian or a Difference-of-Gaussians (DoG) response function would be a better explanation for the data. The DoG function has an excitatory centre and inhibitory surround. The second question asked what shape of receptive field would best explain the data (circular, elliptical or elliptical with rotation).

Thus, we formed a 2x3 factorial model space to address these questions. Here we describe the modifications made to the basic circular (isotropic) model to form these alternative models.

**Elliptical models**

We specified models which replaced the pRF width (standard deviation) parameter $\sigma$ with separate parameters representing horizontal width $\sigma_x$ and vertical width $\sigma_y$. This enabled the pRF to take on an elliptical shape. This was implemented by changing the covariance matrix $\Sigma$ (see Equation 2) to include two separate parameters:

$$\Sigma = \begin{bmatrix} \sigma_x^2 & 0 \\ 0 & \sigma_y^2 \end{bmatrix} \quad (16$$

Latent variables were introduced for each of the two width parameters, as for the single width parameter in the basic model, with the same priors.

**Elliptical models with rotation**

Enabling the pRF to rotate required the introduction of a parameter P representing the correlation coefficient:

$$\sigma_P = P * \sigma_x * \sigma_y \quad (17$$

$$\Sigma = \begin{bmatrix} \sigma_x^2 & \sigma_P \\ \sigma_P & \sigma_y^2 \end{bmatrix}$$

Parameter P was constrained within the range -1 to 1 by substituting it for a latent variable $l_P$. The latent variable was transformed within the model as follows:

$$P = 2 * NCDF(l_P, 0, 1) - 1 \quad (18$$

**Difference of Gaussians**

We created versions of the circular, elliptical and elliptical + rotation models which used a Difference of Gaussians response function (Rodieck, 1965; Zuiderbaan et al., 2012). The existing Gaussian distribution was complemented by a second distribution, which had certain constraints to ensure that it would act as an inhibitory surround. For the basic case of a circular receptive field, the modified response function was as follows:

$$z(t) = \beta_c \sum_{p_i \in U_t} N(p_i | \mu, \Sigma_c) - \beta_s \sum_{p_i \in U_t} N(p_i | \mu, \Sigma_s)$$

$$\Sigma_c = \begin{bmatrix} \sigma^2 & 0 \\ 0 & \sigma^2 \end{bmatrix}$$

$$\Sigma_s = \begin{bmatrix} (\sigma^2 + \sigma_d^2) & 0 \\ 0 & (\sigma^2 + \sigma_d^2) \end{bmatrix} \quad (19$$

$$\beta_s = \max(\beta_c - \beta_d, 0)$$

$$(\beta_c, \beta_d, \sigma_d) > 0$$

Here, the neuronal response $z$ was the difference of two normal distributions. The first represented the excitatory response of the centre of the pRF. It was parameterised by scaling parameter $\beta_c$, centre location $\mu$ and covariance matrix $\Sigma_c$. The second normal distribution represented the inhibitory surround of the pRF. It has the same location as the excitatory centre $\mu$, but different scaling $\beta_s$ and covariance $\Sigma_s$. The difference in scaling between the two pRFs was controlled by parameter $\beta_d$ and the difference in the pRFs' widths was controlled by the parameter $\sigma_d$. Positivity constraints were enforced on the $\beta$ scaling parameters and $\sigma_d$. This formulation, which introduced 2 new free parameters compared to the previous model ($\beta_d, \sigma_d$), ensured that the response of the centre was positive and the response of the surround was negative.

Using a latent variable, we constrained parameter $\sigma_d$ to fall into the range $[0 \; r]$ with a flat prior. The difference in scaling of the surround relative to the centre of the pRF, $\beta_s$, was constrained to be positive using a log parameter. This translated to a prior expectation for $\beta_s$ of 0.05 with 95% confidence interval [0.001 1.97].

### 2.9.8 Bayesian model comparison

We compared the evidence for our data under 6 different pRF models, using the free energy, described above. For each voxel we had 6 numbers - the free energy of each competing model - which we collated into a [voxels x 6] matrix. We submitted this matrix for analysis using a random effects (RFX) model (Stephan et al., 2009). Typically, this approach is used in the context of studying groups of subjects. One assumes that each subject's data were generated by one of the models in the comparison, and the RFX model estimates the probability that any randomly selected subject's data were generated by each model. These are known as the expected probabilities of each model. Additionally, it gives the probability that any one model is better than all the others in the comparison – these are the protected exceedance probabilities. (The word 'protected' refers to a statistical correction to account for the possibility that all models are equally likely (Rigoux et al., 2014).) Here, we simply used multiple voxels instead of multiple subjects, and the RFX model identified which pRF best explained the data in spatially extended regions of the brain.

After performing model comparison at the level of all included voxels, we next performed comparisons on a region of interest (ROI) basis. We used anatomical masks supplied with the SPM Anatomy Toolbox version 2.2b (Eickhoff et al., 2005), which were defined based on cytoarchitectonic observations from a series of studies (Amunts et al., 2000; Kujovic et al., 2013; Malikovic et al., 2015; Rottschy et al., 2007). These regions were hOC1, hOC2, hOC3d, hOC4d, hOC3v, hOC4v, hOC4la and hOC4lp. The closest corresponding functionally defined regions are V1, V2, V3d, V3a, V3v, V4v, LO-1 and LO-2 respectively, however this correspondence between cytoarchitecture and function is not exact, nor does it take into account between-subjects variability. We made one modification to the masks, which was to split the hOC2 mask into dorsal and ventral parts to give hOC2d and hOC2v, situated either side of hOC1. We inverse warped each mask from MNI space into the native space of our subject and used these custom masks to constrain our model comparisons.

Strictly, the RFX procedure described above assumes that the sources of data (the voxels) are independent. This assumption was violated here due to the spatial smoothness of the fMRI data, but nonetheless this method allowed us to provide summary measures for each ROI (see section 3.5.2). For completeness, we repeated the pRF analyses and model comparison while removing the issue of spatial smoothness assumptions, by using the mean timeseries for each ROI rather than each individual voxel. We initially found the model fits to be very poor. This was because the prior on observation noise (log precision of 4) was no longer realistic – averaging over voxels had essentially eliminated any zero-mean Gaussian noise (due to central limit theorem). We therefore increased the prior log precision of the noise to 10, effectively stating that in these averaged timeseries, we did not expect there to be any (zero-mean Gaussian) observation noise. The results of this analysis are illustrated in Supplemental Figure 1 and are summarised in Section 3.5.2 of the results.

# 3 Results

## 3.1 Parameter estimates

After estimating a pRF model for each voxel, we produced maps of the parameters (Figure 3). These confirmed that the parameters were in keeping with established properties of visual cortex. There was a gradient of pRF centre position, from the periphery to the fovea, along the midline of occipital cortex (Figure 3, left) and each hemisphere responded to the contralateral visual field (Figure 3, middle). pRF widths were estimated to be smallest in early visual cortex, becoming larger as one descends into temporal cortex (Figure 3, right).

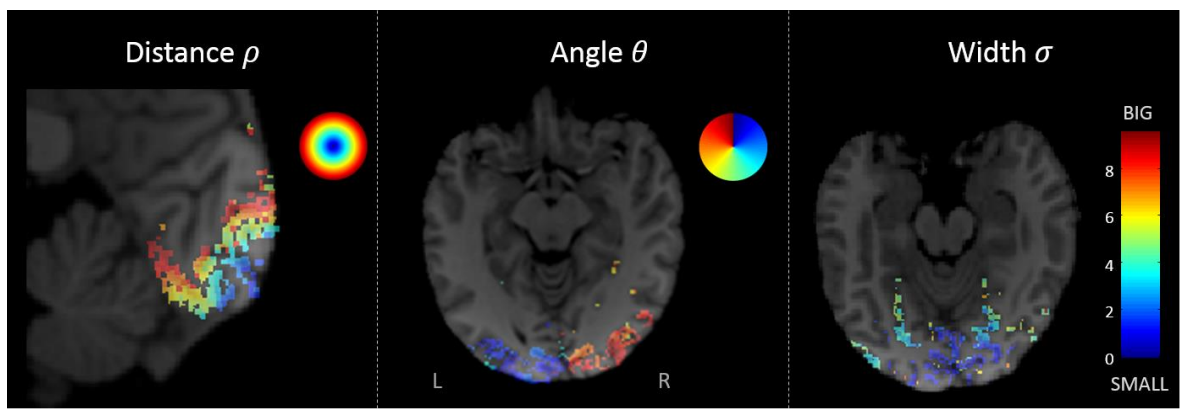

*Figure 3 Parameter maps after estimating the basic pRF model. Left: Distance of the pRF centre from the middle of the visual field (eccentricity), with colours indicating degrees of visual angle, projected onto a sagittal slice of the subject's structural MRI, close to the midline. Middle: Polar angle of the pRF within the visual field, with units of radians, shown in the axial plane. Right: Width parameter (standard deviation), with the colour bar indicating degrees of visual angle - smaller pRFs are in blue, larger in red. Maps thresholded at model posterior probability > 0.95 (see methods Section 2.9.1).*

The remaining parameters controlled the scaling of the neuronal response ($\beta$), the transit time ($\tau$), the rate of decay ($\kappa$) and ratio of intra- to extra-vascular signal ($\epsilon$), all of which were estimated on a voxel-wise basis together with the pRF parameters. To investigate whether estimating the neurovascular / haemodynamic response on a voxel-wise basis was justified – i.e. whether there was variation across voxels – we calculated the estimated BOLD impulse response function for each voxel and summarised these responses by their mean and standard deviation (Figure 4). This revealed considerable variability in the height of the peak across voxels, emphasising the importance of estimating haemodynamic as well as neuronal parameters in the model.

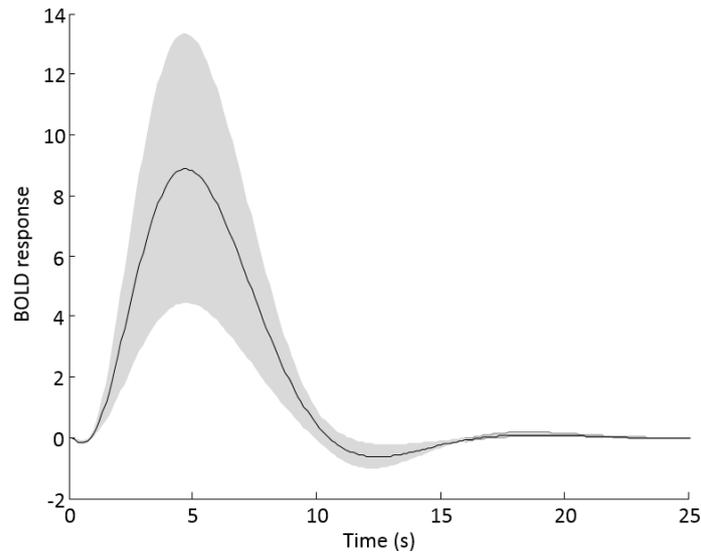

*Figure 4 The variability of the estimated BOLD impulse response function across voxels, in response to a stimulus at time 0 of duration 0. The solid line is the mean across voxels and the shaded area is mean +- 1 standard deviation across voxels.*

The prior response, posterior response and fitted timeseries for an example voxel are shown in Figure 5. In Figure 5A, the response of the pRF model is shown for each point in retinotopic space, with the model's parameters set to their prior means and variances. To account for uncertainty, we sampled from the parameters according to their prior probability distributions, then generated the pRF response based on the sampled parameters. Repeating this 1000 times and averaging the responses gave the prior Predictive Density (PD) shown. Figure 5B shows the response of the model with parameters estimated from the data (the posterior parameters), demonstrating a punctate pRF estimate in the left visual field. This plot again takes into account uncertainty over the parameters through sampling (the posterior PD). Strikingly, the uncertainty visible as the diffuse pattern in the prior PD has been explained away after seeing the data. This close fit to the data can be also be seen qualitatively in the modelled timeseries (Figure 5C). We noted that the post-stimulus undershoot in the BOLD response was not well captured, which is a known limitation of the haemodynamic model we are applying (Havlicek et al., 2015). We return to this issue in the discussion.

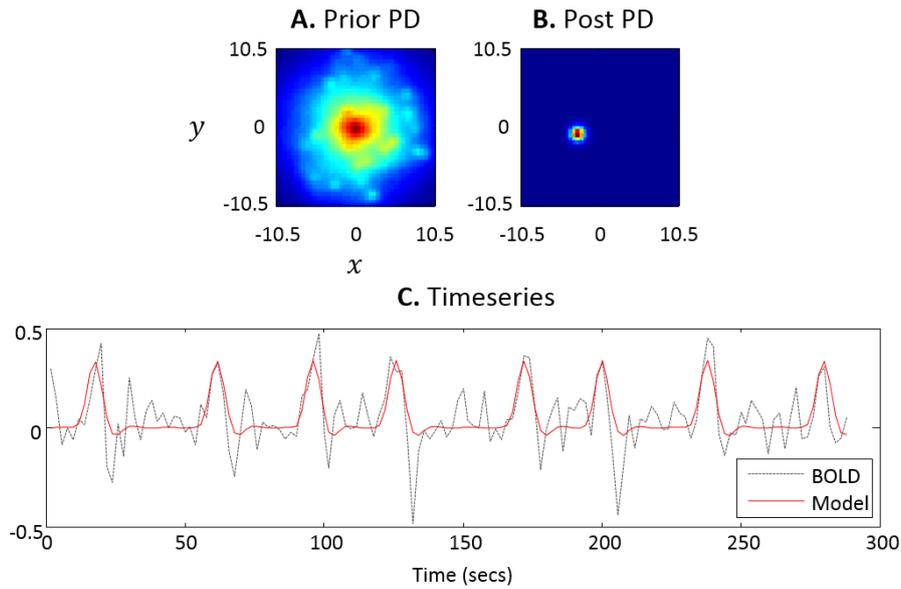

*Figure 5 Estimated pRF and fitted timeseries for an example voxel in putative right V1 (voxel 14381).* ***A.*** *The response of the model to each position in visual space prior to Bayesian model fitting. This is the prior Predictive Density (PD), which uses sampling to take into account uncertainty over the parameters. The prior model parameters were sampled 1000 times and the predicted responses generated and averaged. The units of the axes are degrees of visual angle.* ***B.*** *The posterior PD, which is the estimated pRF response after Bayesian estimation.* ***C.*** *The modelled timeseries (red) and observed timeseries (grey).*

## 3.2 Correlation and entropy

A key advantage of the Bayesian approach is that the uncertainty of each parameter (the variance) is estimated, as is the covariance among parameters. The covariance is important because parameters which strongly covary cannot be precisely estimated, and this uncertainty should be taken into account when making inferences. To investigate the covariance among parameters, we averaged models over voxels using Bayesian Parameter Averaging (see methods Section 2.9.4) and then transformed the averaged covariance matrix to give a correlation matrix. This is shown in Figure 6, where the upper left quadrant shows the correlations among neuronal (pRF) parameters, the bottom right quadrant shows correlations among haemodynamic parameters and the top right quadrant shows the correlation between neuronal and haemodynamic parameters.

Reassuringly, the parameters representing the location of the pRF in visual space $(l_\rho, l_\theta)$ had only weak correlation with other parameters. However, the pRF width parameter $\sigma$ was strongly correlated with the scaling parameter $l_\beta$ (Pearson's correlation 0.63). In practice, this means that one could increase the pRF width or increase the scaling parameter and, to some extent, get the same change in the predicted timeseries. This phenomenon is not specific to the Bayesian approach proposed here, and may be of concern for any pRF estimation method, despite this correlation not generally being quantified. The correlation matrix also revealed statistical dependencies among the haemodynamic parameters, as well as between the scaling parameter $l_\beta$ and the haemodynamic parameters. In practice, correlations among the haemodynamic parameters are unlikely to be a problem, as experimenters rarely need to make inferences about these parameters individually.

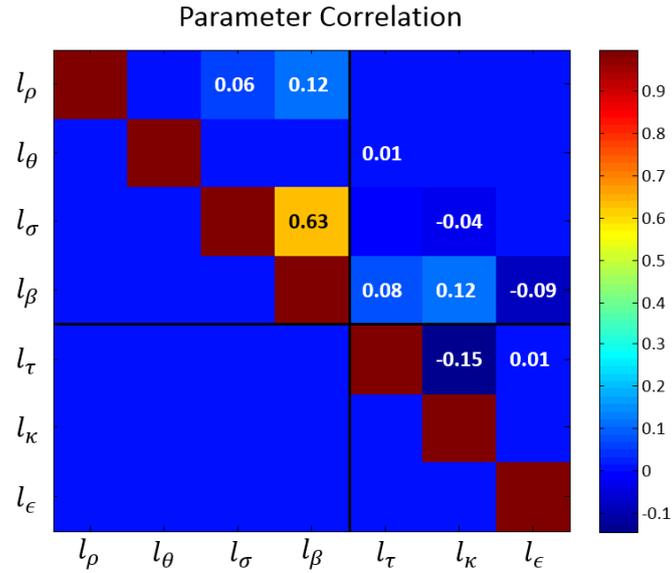

*Figure 6 Correlation among model parameters (latent variables), averaged over voxels. There were 4 neuronal (pRF) parameters ($\rho, \theta, \sigma, \beta$) and 3 neuro-vascular / haemodynamic parameters ($\tau, \kappa, \epsilon$). Only values with an absolute value greater than 0.01 are labelled.*

Having covariance among parameters induces uncertainty in the parameter estimates, in addition to any uncertainty induced by observation noise. Next, we mapped this uncertainty by computing the entropy for each pRF model, where more positive values indicate greater uncertainty about the parameter estimates. We computed separate entropy maps for the pRF location ($l_\rho, l_\theta$) and width ($l_\sigma$) parameters. The entropy map for pRF location (Figure 7, top) showed that we could be most certain about the position of the pRF in early visual cortex, and as one moves laterally or descends into temporal cortex, the uncertainty increases. There was less confidence overall for pRF width than for pRF location – the minimum entropy was -5.75 nats for pRF width compared to -18.99 nats for pRF location (lower is more confident). Surprisingly, in early visual cortex - where we would expected to have been most certain about pRF width - we were in fact least certain (Figure 7, bottom). This could be because smaller pRFs, which typify early visual cortex, require higher resolution stimuli in order to gain confident estimates about their size. Additionally, uncertainty over pRF width may have been induced by covariance with the scaling parameter $l_\beta$, identified above.

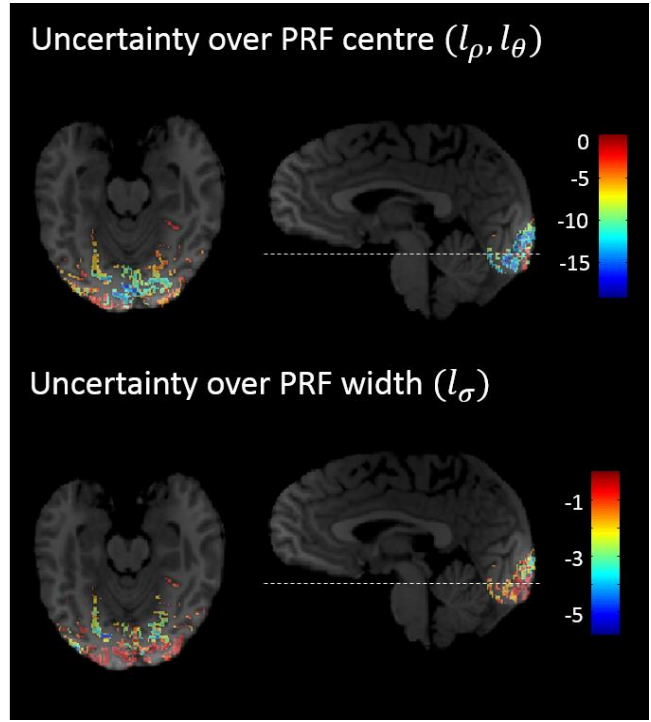

*Figure 7 Entropy maps showing uncertainty over the pRF location (top) and uncertainty over the pRF width (bottom) after estimation of the models. More positive values (red) indicate greater uncertainty. Colours indicate entropy in units of nats.*

## 3.3 Face validity

We next performed two forms of validation: we used simulations to establish the face validity of the approach and we used empirical data to evaluate test-retest reliability across runs.

We fitted pRF models to 1000 simulated timeseries where the 'true' parameters which generated the data were known, under varying levels of observation noise. The expected values of the pRF position, width and neuronal scaling parameters closely matched the true parameters (Figure 8, left), with the width and beta parameters only dropping markedly in accuracy below an SNR of 0.75. While all correlations were highly significant (all p < 4.96e-4), the neurovascular / haemodynamic parameters ($\tau, \kappa, \epsilon$) were estimated less accurately and were more sensitive to the SNR than the pRF parameters. This may have been due to the covariance among the haemodynamic parameters, reducing the identifiability of their expected values. The accuracy of the $\theta$ polar angle parameter was slightly lower than we had expected (correlation 0.96 at maximum SNR). We found this was an artefact of the use of polar coordinates. There were pRFs with parameter $\theta$ close to $-\pi$ radians in the test set and close to $+\pi$ radians in the validation set, or vice versa. Despite being adjacent in visual space, the difference in sign reduced the test-reliability. When we corrected for this by converting from polar to Cartesian coordinates, accuracy of the $\theta$ parameter was at ceiling (correlation 1.00) for SNR 1.56, SNR 1.25 and SNR 1, reducing to correlation 0.99 at SNR 0.75 and correlation 0.97 at SNR 0.5.

We also investigated the accuracy of the estimated uncertainty of the parameters (Bayesian confidence intervals). We anticipated that the posterior 95% confidence interval of each parameter would include the "true" parameter which generated the simulated data, approximately 95% of the time. For each level of SNR and for each parameter, we computed the fraction of the 1000 voxels which included the 'true' parameter in their estimated 95% confidence interval (Figure 8B). At the estimated SNR of our data (SNR 1.56), the confidence intervals behaved as expected. With reducing SNR, the accuracy of the confidence intervals decayed slowly (in particular the width parameter), perhaps due to the appearance of local optima in the parameter space, coupled with the correlation among certain parameters. However, overall, these results demonstrate that our estimates of uncertainty were robust to noise, even for the haemodynamic parameters which had the least accurate expected values.

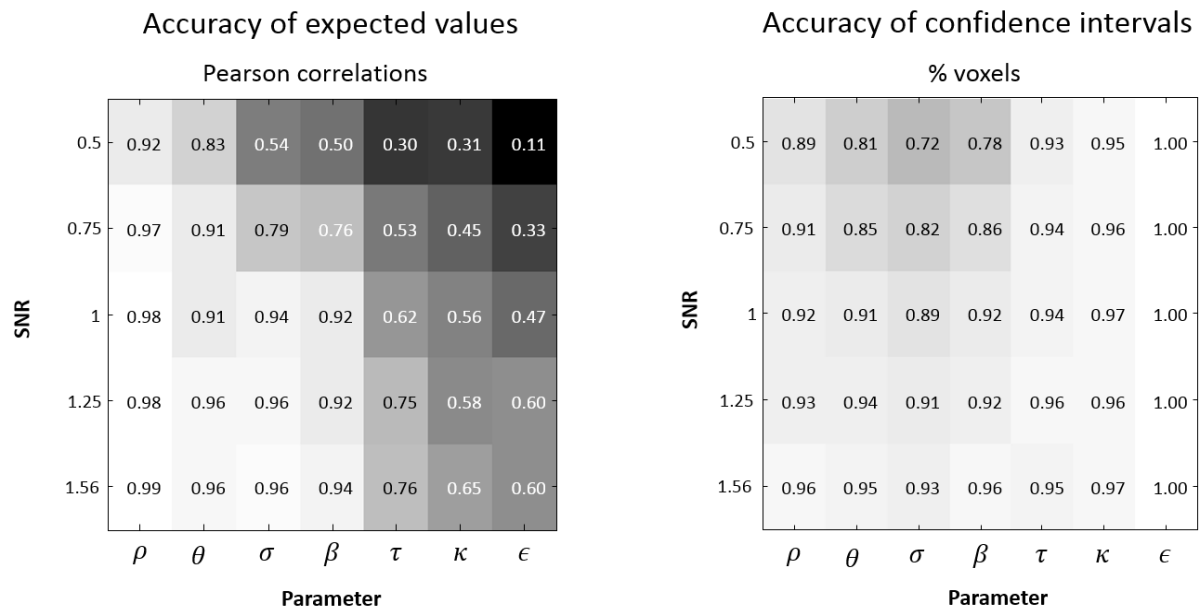

*Figure 8 Accuracy of parameter estimates assessed by simulation.* **Left:** *Pearson's correlations between the parameters used to simulate data and parameters estimated from those simulated data. The vertical axis lists 5 levels of observation noise, quantified as the signal-to-noise ratio (SNR). Darker colours indicate lower correlations. Parameters as for Figure 6.* **Right:** *The percentage of voxels where the estimated 95% (posterior) confidence interval contained the parameter which generated the data.*

## 3.4 Test-retest reliability

Next we tested the consistency of the parameter estimates across scanning runs on empirical data. The main analysis, above, used only the 8 odd-numbered runs. Here we used the 8 even-numbered runs as a validation set, correlating parameter estimates from each half of the data on a voxel-wise basis. Note that this made a strong assumption about the success of coregistration – it assumed that the voxels in all runs were perfectly aligned. Nevertheless, the correlation between parameter estimates from odd- and even-numbered runs was strong (Table 1). Converting from polar coordinates $(\rho, \theta)$ to Cartesian coordinates $(\mu_x, \mu_y)$ raised the test-retest accuracy to 0.98 and 0.95 for the $\mu_x$ and $\mu_y$ coordinates respectively.

**Table 1: Correlations between parameters estimated from alternating runs**

|   | Pearson's correlation |
|---|---|
| $\rho$ | 0.92 |
| $\theta$ | 0.89 |
| $\sigma$ | 0.85 |
| $\beta$ | 0.90 |
| $\tau$ | 0.89 |
| $\kappa$ | 0.88 |
| $\epsilon$ | 0.83 |

## 3.5 Model comparison

Having validated the modelling framework, we next used it to ask two questions. First, we asked whether the neuronal population within each voxel had only an excitatory response to stimuli within their receptive fields, or whether they also exhibited an inhibitory response at the periphery of their receptive fields. We represented these two alternatives using models with a single Gaussian response function, and models with a Difference of Gaussians (DoG) response function with an excitatory centre and inhibitory surround, as has previously been introduced (Zuiderbaan et al., 2012). We also asked which of three shapes of receptive field would offer the best explanation for our data – circular (Model 1), elliptical (Model 2), or elliptical with rotation (Model 3). We formally addressed these questions by comparing the evidence for the data under competing models (Bayesian model comparison). The model that wins in such a comparison is the model that strikes the optimal balance between accuracy and complexity.

We formed a model space akin to a 2x3 factorial design. The first factor was the response function: models either had a single Gaussian or a DoG response function. The second factor was the shape of the receptive field (x3). We specified each of these 6 models and estimated the free energy (log model evidence) within each voxel. Thus, we had six free energies per voxel, representing the relative evidence for each model, which we pooled across voxels using a random effects analysis (see methods section 2.9.8 for details).

### 3.5.1 Whole volume analysis

Figure 9A shows the expected probability of each model. If a voxel were picked at random, these are the probabilities of each model having generated that data. The model with the strongest probability was the DoG Model 1 – that is, a model with circular receptive field and excitatory / inhibitory dynamics. Thus, if choosing a voxel at random, there would be a 55% chance that DoG model 1 would be the best model, $p(m|y) = 0.55$. The next best model was the single Gaussian Model 1, with $p(m|y) = 0.29$. The remaining models all had much weaker probability. Figure 9B shows the Protected Exceedance Probability (PXP), which is the probability that any one model is better than all other models in the comparison. With probability approaching certainty, DoG Model 1 was the winning model, and therefore the best explanation for any randomly chosen voxel's data. An example timeseries fitted by DoG Model 1 is shown in Figure 9C. Together, this comparison demonstrates that DoG is a better explanation than a single Gaussian model overall, and having elliptical and angled pRFs does not, at the level of the whole visual cortex, contribute enough to the model evidence to outweigh the added complexity they induce.

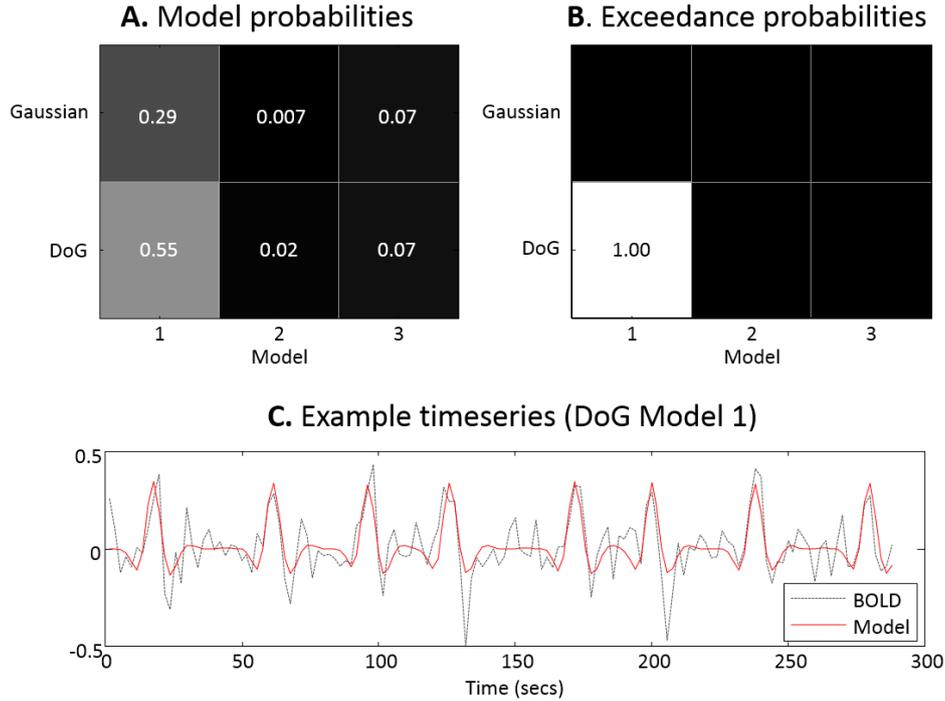

*Figure 9 Comparison of 6 pRF models across all included voxels. Models either had a single Gaussian or Difference of Gaussians (DoG) receptive field (rows) and were circular (Model 1), elliptical (Model 2) or elliptical with rotation (Model 3), shown in the columns. **A**. The expected probability of each model. **B**. The protected exceedance probability (PXP) of each model, which is the probability that each model is better than the all the others in the comparison. **C**. An example timeseries for the winning model (DoG Model 1), fitted to data from the same voxel as shown in Figure 5.*

### 3.5.2 Regions of interest analysis

We next investigated whether the 6 competing pRF models had different explanatory power in different brain regions. We repeated the model comparison within each of 9 regions of interest, based on predefined (cytoarchitecturally defined) anatomical masks.

In most regions (hOC1, hOC2d, left hOC2v, hOC3d, hOC3v, hOC4lp), a DoG model rather than a single Gaussian model best explained the data (Figure 10, black grids). The single Gaussian models were preferred in right hOC2v, hOC4d, right hOC4v and hOC4la. The result in left hOC4v was ambiguous, with $xp = 0.36$ for Gaussian Model 1 and $xp = 0.64$ for DoG Model 1, which may indicate a heterogeneous population of response in this region. So with regards to our first question of whether there was evidence for an excitatory-inhibitory receptive field, we can conclude that this was the case in most regions lower in the visual hierarchy, however the benefit was not seen in the majority of the highest level regions we modelled, where a simpler Gaussian model was selected as the winner.

Our second question regarded the shape of the receptive field. In most regions (hOC1, hOC2d, hOC3d, hOC4d, hOC2v, hOC3v, left hOC4v, hOC4lp), the model with a simple circular receptive field (Model 1) was the winner. Therefore, modelling an elliptical shape and rotation did not improve the model fit sufficiently to outweigh the added complexity. The clear exception was bilateral hOC4la, where there was strong evidence (PXP > 0.99) that the single Gaussian angled model (Model 3) was the winner. In right hOC4v, the evidence was divided between the Model 1 and Model 3.

Having identified the best model in each region, we next visualised the extent of the visual field covered by these models' receptive fields. For each ROI, we summed the (posterior) receptive field of the overall winning model across voxels, to form coverage maps (Figure 10, right columns). This confirmed that each pRF responded primarily to the contralateral side of retinotopic space, with dorsal regions responding to the lower visual field and ventral regions responding to the upper visual field. Lateral regions responded preferentially to the lower visual field, and strikingly, the coverage area of the bilateral hOC4la models was angled towards the fovea. To quantify the angle estimates in hOC4la, we averaged the parameters across voxels using Bayesian Parameter Averaging. This gave a probability distribution representing the average correlation coefficient (rotation parameter), with an expected value of 0.08 in left hemisphere and -0.37 right hemisphere. The probability that these parameters were non-zero were 0.99 and 1.00 respectively.

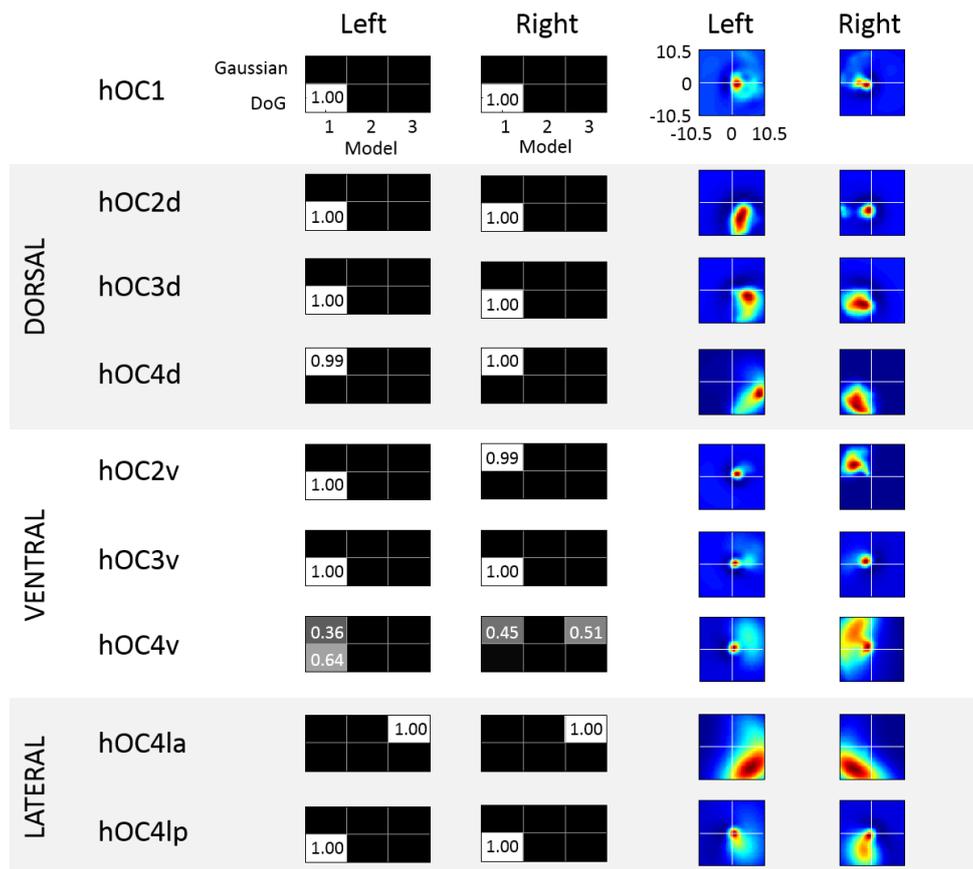

*Figure 10 Results of comparing 6 different pRF models (black grids in the left 2 columns) in 9 regions of interest. **Left columns**: Each grid shows a 2x3 factorial model space. The models either employed a 2D Gaussian distribution or a Difference of Gaussians (DoG). For each type there were three variants: Circular (Model 1), Elliptical (Model 2) or Elliptical with Rotation (Model 3). The shades of grey / white and labels indicate the Protected Exceedance Probability (PXP), which is the probability that a model is better than all the others being tested. **Right columns:** The summed pRFs of the winning model within each region (see text for details). Colours are scaled within each plot individually, representing arbitrary units of neuronal activity. Region identities are detailed in Section 2.9.8.*

Finally, we performed an analysis to investigate whether similar results would be found by fitting pRF models to the mean timeseries of each region, rather than fitting a pRF model to every voxel independently (Supplementary Figure 1). Note that the model which best explains the average response for an extended region is not necessarily the model which best explains individual voxels' timeseries, due to variability across voxels in pRF response (Figure 3). Nonetheless, in this analysis using far less data, every region with the exception of left hOC1 favoured a DoG rather than a Gaussian response function. Whereas individual voxels' receptive fields were best described with a circular model (DoG Model 1, Figure 10), the averaged timeseries in most regions were best explained with an angled and rotated receptive field (DoG Model 3). The pRF response functions in most regions were remarkably similar to the summed voxel-level response functions shown in Figure 10.

# 4 Discussion

In this paper we introduced a Bayesian framework for specifying, estimating and comparing pRF models. The framework is generic and so can be applied to stimulus spaces of any dimension; here we evaluated it for mapping 2D visual receptive fields. Having validated the method using both simulations and a simple evaluation of test-retest reliability, we demonstrated how it may be used to test hypotheses about neuronal response functions, at the voxel level and within predefined regions of interest.

A key advantage of this approach is that the uncertainty of each parameter estimate is estimated, by computing the full covariance matrix of the parameters. This gave rise to a striking observation – there was a strong correlation between the neuronal scaling parameter $l_\beta$ and the pRF width parameter $l_\sigma$. In other words, increasing either parameter would, to a certain extent, give the same change in the modelled BOLD response, which limits the precision with which either can be estimated. An appropriate way of dealing with this is to take account of the uncertainty when making inferences. Here, we did this by plotting the pRF's posterior predictive density (PD) to visualise the receptive field, which represents the uncertainty in the pRF position and width. This made clear that with the relatively high SNR data we used, which had been scanned at 7T and averaged over 8 runs, the receptive field was estimated with high precision despite the covariance between parameters. It will be interesting to see how the precision of the parameters vary when acquiring empirical data at lower field strengths. As suggested by our simulations, with reduced SNR, taking uncertainty into account becomes increasingly important.

In the generative model implemented here, we included an established model of neurovascular coupling and BOLD response. Whereas most pRF studies use a canonical haemodynamic response function, which may be tailored on a subject-by-subject basis, we estimated parameters of the haemodynamic model simultaneously with the pRF parameters, on a voxel-wise basis. (Readers familiar with Dynamic Causal Modelling will note that we are effectively estimating a single region DCM for every voxel.) This comes with a computational cost – estimation takes longer than standard approaches to pRF estimation. However, our results demonstrate the utility of this voxel-wise model. There was considerable variability in the peak (and to a lesser degree, the undershoot) of the BOLD response across voxels (Figure 4), which would not be modelled using a fixed BOLD response across voxels. Reassuringly, the parameters of the haemodynamic model were only weakly correlated with the parameters representing the pRF location and width (Figure 6), meaning neuronal and haemodynamic contributions to the BOLD timeseries can be separated with high confidence.

We demonstrated testing hypotheses about pRF response function or shape by embodying each hypothesis as a model and comparing the evidence for each. It has previously been stated that "pRF models…are not centered on statistical hypothesis testing…Rather, the strategy follows the constructivist philosophy of creating models to account for an increasingly large range of stimuli in many visual areas." (Wandell and Winawer, 2015, p. 350). Developing pRF models that are as general as possible requires identifying those which offer the best trade-off between accuracy and complexity. A penalty for complexity (also called an Occam factor) is required when comparing models, otherwise the most complex model will always be considered the best (explain the most variance), at the risk of overfitting the noise and failing to generalise. Finding the best of several models in terms of balancing accuracy and complexity is achieved by identifying the model with the highest model evidence, a procedure referred to as Bayesian model comparison. Here we used an approximation of the model evidence, the free energy, to score models. Unlike other approximations such as the AIC and BIC, the free energy takes into account the full covariance among parameters (Penny, 2012), which reflects the fact that independent parameters contribute more to a model's complexity than correlated parameters.

Using Bayesian model comparison, we first asked whether the response in each voxel is best described as a single multivariate normal distribution, or as a Difference of Gaussians (DoG) model with excitatory centre and inhibitory surround (Rodieck, 1965; Zuiderbaan et al., 2012). The DoG model had previously been found to explain more variance in V1, V2 and V3, but offered little or no improvement in higher level regions V3a, hV4 and LO1 (Zuiderbaan et al., 2012). Similarly, using Bayesian model comparison, we found DoG models to be better than single Gaussian models in cytoarchitecturally defined regions roughly corresponding to V1, V2 and V3, with the exception of right hOC2v, where there was strong evidence for the basic single Gaussian model. In higher level regions, we found strong evidence for the simpler single Gaussian model. This was the case in region hOC4d (which may mostly closely align with V3a of Zuiderbaan et al. (2012)) and region hOC4la (which may align with LO-1). The fact that we found strong evidence for a simpler model in higher regions, and Zuiderbaan et al. (2012) found little or no difference in explained variance in these higher regions, speaks to the notion that the model evidence will favour the simpler model unless there are sufficient gains in accuracy to outweigh the added complexity.

We found that the single Gaussian pRF model, coupled with the Balloon model of BOLD response, failed to capture the majority of the post-stimulus undershoot in the BOLD response (Figure 5C). These parts of the timeseries appear to have fitted slightly better using a DoG model, and may coincide with the inhibitory activity in centre-surround dynamics (Figure 9C), although a noticeable post-stimulus undershoot remains unmodelled. A recently proposed upgrade to the BOLD response model we used here offers a more physiologically plausible representation of the post-stimulus undershoot (Havlicek et al., 2015), which could be included in our framework. Model comparisons could then be performed to ensure that there is still strong evidence for a DoG neuronal response function, in the context of a haemodynamic model which better explains the BOLD undershoot.

Having identified the DoG model as being preferred in most regions, our second question regarded the shape of the receptive field. We compared three receptive field shapes – circular, elliptical and elliptical with rotation – and tested the evidence for models with these specifications in sub-regions of the brain. In most regions, a simple circular model was the winner - meaning the data could not support a more complex explanation involving an elliptical shape or rotation. We could only be confident that a more complex model was the winner in bilateral V4la. Coverage maps (Figure 10, right) showed that the pRFs in these regions covered an area of the contralateral lower visual field, which were angled towards the fovea. While anatomically defined region V4la was proposed to correspond to functionally defined region LO-1 (Malikovic et al., 2015), it may be more closely related to region of V3B of Smith et al (Smith et al., 1998), which was defined as responding to the contralateral lower hemifield. For a detailed discussion on the definitions of regions within LOC, we refer readers to Larsson and Heeger (2006).

We do not wish to draw strong conclusions from these region-of-interest analyses, as we were only using data from a single subject, and we did not use functionally-defined regions demarcated on the cortical surface. Indeed, these results may suggest limitations to the correspondence between anatomically defined and retinotopically defined regions. For instance, hOC4d has been suggested to align with functional region V3a (Kujovic et al., 2013). However, V3a has a full hemifield representation, whereas in our results we only found coverage of the lower hemifield. These uncertainties will be resolved by extending this modelling framework to operate at the cortical surface, so that subject-specific visual regions can be delineated. With the analyses presented here, we simply hope to have demonstrated that pRF models can be formerly compared based on their model evidence and summarised within regions of interest, to address interesting hypotheses. We found that, of the models we tested, the most parsimonious (simplest) model that explained the most data was a Difference of Gaussians model with a circular (isotropic) receptive field (Figure 9).

A limitation on the explanatory power of the pRF model used here, and of pRF models more generally, is that they are phenomenological. They are very effective at describing the summed receptive field of a voxel or a brain region, but they give no insight into how this receptive field arises from the underlying neuronal circuitry. Developments in modelling neuronal circuitry using functional imaging data, Dynamic Causal Modelling (Friston, 2003), have come about by iteratively developing models which afford greater model evidence than their predecessors. By providing a framework for evaluating models, we hope to facilitate the development of biologically plausible generative models for pRFs. These are likely to be spatiotemporal models which explain how lateral connections between neuronal populations give rise to the distributed pattern of activity across the cortical sheet. Neural field models (Moran et al., 2013; Pinotsis and Friston, 2011), which have so far been applied in the context of magneto- and electro-encephalography recordings, offer a first step in this direction.

Future work will be in several directions. Estimation speed of the pRF models was generally over 100 seconds per voxel (around 2 hours, 40 minutes in total on the 192-core cluster computer we used to estimate 14,395 voxels). As such, we expect this framework to be primarily used in a voxel-wise manner for small regions of interest, or alternatively on summary timeseries (the mean or first principal component) from regions of interest, although the latter is not guaranteed to identical results (Supplementary Figure 1). However, there may be opportunities for software optimisation, for instancing removing the overhead of converting between polar and Cartesian coordinates. In the meantime, we have addressed the performance issue by providing software tools for distributed model estimation over parallel processors (Appendix A), and in future this could be extended to take advantage of massively parallel GPU computing, which has already been demonstrated in the context of Dynamic Causal Modelling (Aponte et al., 2016). It would also be helpful to develop integration with the Freesurfer software package, to enable projection of results to the cortical surface. In terms of applications, the main focus of our ongoing efforts will be to model higher dimensional spaces. The generality of the approach makes it ideally suited to mapping multimodal or abstract stimulus spaces onto the brain, with any number of model parameters. We hope that the tools we have implemented (Appendix A) will be useful to researchers in a variety of fields.

# 5    Appendix A: Software implementation

We have developed a suite of software tools for specifying, estimating and comparing pRF models. These are Matlab functions which depend on the open-source SPM software package (http://www.fil.ion.ucl.ac.uk/spm/), and use the same underlying algorithms as Dynamic Causal Modelling (Friston, 2003). This toolbox forms the reference implementation of the Bayesian framework we have described, although we hope to have given sufficient detail in the manuscript that it could be re-implemented in other software packages. The BayespRF toolbox is available from https://github.com/pzeidman/BayespRF .

The toolbox is structured around a series of operators which act upon pRF models. A pRF model, together with one or more timeseries, is represented as a Matlab structure and is stored in a file with the format PRF_<name>.mat. A typical workflow proceeds as follows. One performs an initial analysis using a general linear model (GLM) to identify voxels to take forward for pRF analysis (Figure 2). Timeseries are then extracted using the 'Eigenvariate' button in SPM, which also performs all the necessary pre-processing. The resulting file (VOI_<name>.mat) contains both a summary timeseries of all included voxels, and individual timeseries for each voxel. Next, one specifies a pRF model using **spm_prf_analyse('specify',…)**, which provides options for whether to use a summary timeseries or individual voxels' timeseries. If averaging over runs is required, this will be performed at this stage. The model evidence and parameters may then be estimated using **spm_prf_analyse('estimate',…)**. The results of the estimation are reviewed and projected onto 3D orthogonal projections of the brain using the GUI **spm_prf_review(…)**, which will create parameter maps (as shown in Figure 3) if they do not already exist.

If there are a large number of voxels to be estimated, it may be more tractable to use parallel computing and estimate pRF models for multiple voxels simultaneously. The toolbox provides two methods for this. The estimation function has an option 'useparfor', which will take advantage of the Mathworks Parallel Computing Toolbox, if available. Alternatively, a function is provided to take a multi-voxel pRF file and split it into multiple pRF files: **spm_prf_analyse('split',…)**. These individual pRF files can be estimated on separate machines, for instance on separate nodes of a cluster computer, and then merged once estimation is complete using **spm_prf_analyse('merge',…)**. The results of the merged pRF file may then be reviewed.

Within the pRF model file are stored the priors, posteriors, timeseries data and model configuration. A summary of the most important fields are given in Table A1. Users wishing to design new pRF models, or modify the current ones, may wish to take advantage of the **spm_prf_editor** tool, which provides a GUI for manipulating the parameters of a model.

**Table A1: Key fields within the Matlab structure representing a pRF model**

| Field | Description |
| --- | --- |
| M.pE | Prior mean (expectation) of each parameter |
| M.pC | Prior covariance of the parameters |
| M.IS | Name of the pRF model function to use |
| Y.y | Matrix of timeseries data against which to estimate the model |
| xY.XYZmm | Coordinates (mm) of each timeseries in the brain |
| U | [1 x T] structure representing the T stimuli displayed during scanning, including onsets, durations and coordinates of stimulated pixels |
| Ep | Estimated parameters for each voxel |
| Cp | Estimated covariance of the parameters for each voxel |
| Eh | Estimated log precision of the noise |
| F | Variational free energy (approximate evidence) of the model for each voxel |

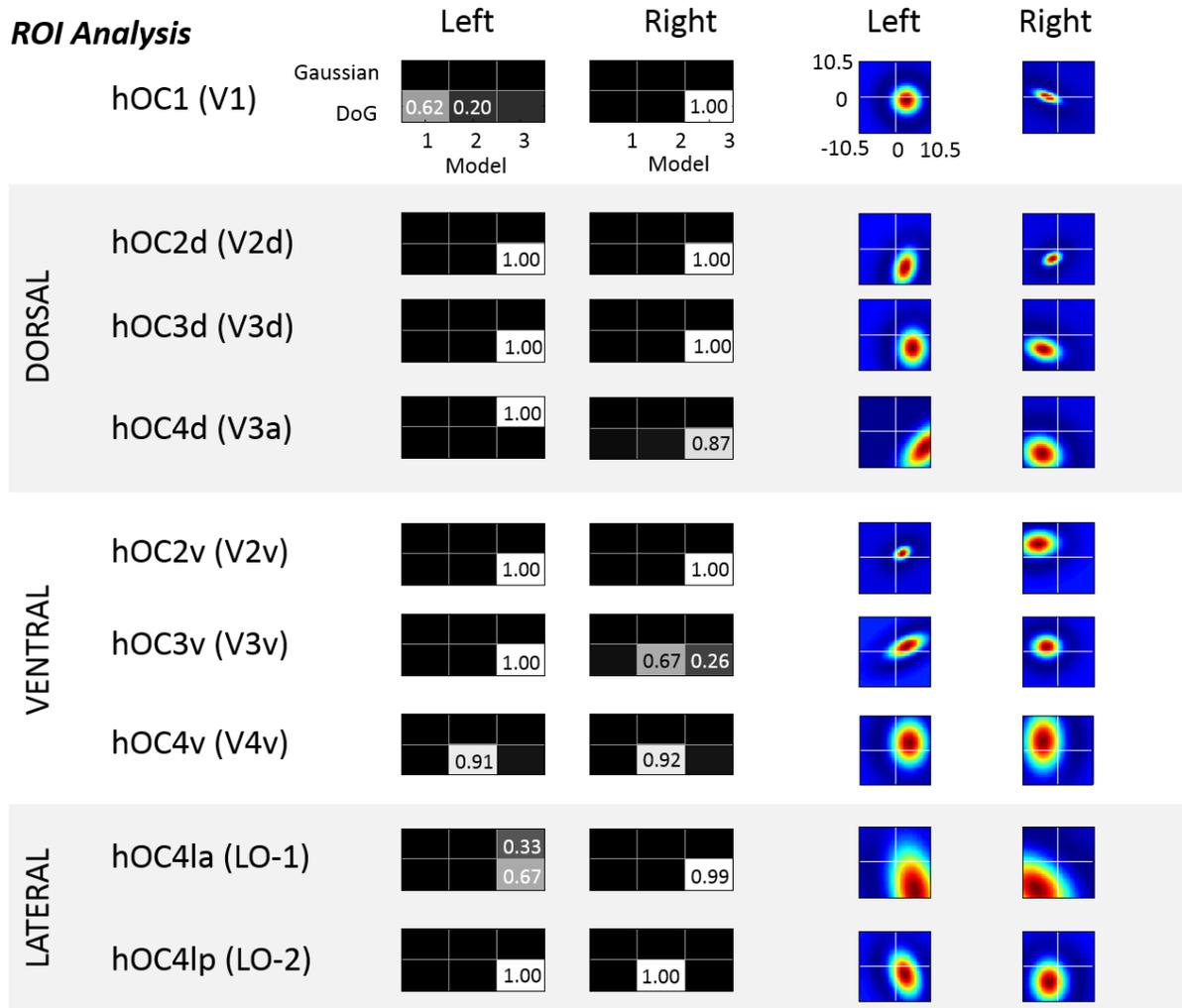

*Figure S1.* Results of comparing 6 different pRF models (black grids in the left 2 columns) in 9 regions of interest, where the data were the mean timecourse from each region. Regions are given their cytoarchitectonic name with the name of closest matching functionally defined region in brackets. **Left columns**: Each grid shows a 2x3 factorial model space. The models either employed a 2D Gaussian distribution or a Difference of Gaussians (DoG). For each type there were three variants: Circular (Model 1), Elliptical (Model 2) or Elliptical with Rotation (Model 3). The shades of grey / white and labels indicate the posterior probability of each model. **Right columns:** The modelled response of the winning model within each region. Colours are scaled within each plot individually, representing arbitrary units of neuronal activity.